\documentclass[letterpaper, 10 pt, journal, twoside]{ieee/IEEEtran}

\usepackage[T1]{fontenc}
\usepackage{amsmath,amssymb} 
\usepackage{url} 
\usepackage{mathtools} 
\usepackage{stfloats} 
\usepackage{wasysym}

\usepackage[dvipsnames]{xcolor} 
\usepackage[caption=false,font=footnotesize,labelfont=footnotesize]{subfig}
\usepackage{xfrac} 
\usepackage{hyperref} 
\usepackage{cleveref} 
\usepackage{siunitx} 
\usepackage{tikz}
\usepackage{tikz-3dplot}
\usetikzlibrary{calc, 3d, angles, decorations.text, fit, quotes, external} 

\crefname{figure}{Fig.}{Figs.}
\Crefname{figure}{Fig.}{Figs.}
\crefname{appendix}{Appx.}{Appxs.}
\Crefname{appendix}{Appx.}{Appxs.}
\crefname{equation}{Eq.}{Eqs.}
\Crefname{equation}{Eq.}{Eqs.}
\crefname{chapter}{Chp.}{Chps.}
\Crefname{chapter}{Chp.}{Chps.}
\crefname{section}{Sec.}{Secs.}
\Crefname{section}{Sec.}{Secs.}
\crefname{table}{Tab.}{Tabs.}
\Crefname{table}{Tab.}{Tabs.}

\newcommand{\papercodeurl}{\url{https://github.com/NeelakshSingh/oct_levitation.git}}
\newcommand{\papervidurl}{\url{https://youtu.be/ZUB8kCYTOmk}}

\newcommand{\vect}[1]{\ensuremath{\boldsymbol{#1}}}       
\newcommand{\mat}[1]{\ensuremath{\mathbf{#1}}}        
\newcommand{\matsubscript}[2]{\ensuremath{\mat{#1}_{#2}}} 
\newcommand{\skewmat}[1]{\ensuremath{\left[ #1 \right]_\times}} 
\newcommand{\specialmat}[1]{\ensuremath{\boldsymbol{\mathcal{#1}}}} 
\newcommand{\transpose}[1]{\ensuremath{#1^\top}} 
\newcommand{\framename}[1]{\ensuremath{\mathrm{#1}}} 
\newcommand{\coordframe}[1]{\ensuremath{\left\{ \framename{#1} \right\}}} 
\newcommand{\func}[2]{\ensuremath{#1\mathopen{}\left(#2\right)}} 
\newcommand{\dfunc}[2]{\ensuremath{#1\mathopen{}\left[#2\right]}} 
\newcommand{\funcdef}[3]{\ensuremath{#1: #2 \to #3}} 
\newcommand{\framevect}[2]{\ensuremath{{}^{\framename{#2}}\vect{#1}}} 
\newcommand{\framevectcomp}[3]{\ensuremath{{}^{\framename{#2}}#1_{#3}}} 
\newcommand{\framevectlabelled}[3]{\ensuremath{{}^{\framename{#2}}\vect{#1}_{#3}}} 

\newcommand{\diag}[1]{\ensuremath{\func{\operatorname{diag}}{#1}}} 

\newcommand{\partialderivative}[2]{
  \ensuremath{
    \mathchoice
      {\dfrac{\partial #1}{\partial #2}}
      {\sfrac{\partial #1}{\partial #2}}
      {\sfrac{\partial #1}{\partial #2}}
      {\sfrac{\partial #1}{\partial #2}}
  }
}

\newcommand{\differential}[1]{\mathrm{d}#1} 

\newcommand{\pseudoinverse}[1]{\ensuremath{#1^{\dagger}}} 

\newcommand{\tderiv}[1]{\ensuremath{\dot{#1}}}         

\newcommand{\varWorldFrameName}{V}      
\newcommand{\varBodyFrameName}{B}      
\newcommand{\varBodyFrame}{\ensuremath{\coordframe{\varBodyFrameName}}}        
\newcommand{\varWorldFrame}{\ensuremath{\coordframe{\varWorldFrameName}}}        

\newcommand{\varMagDipMoment}{\ensuremath{\framevect{\tilde{m}}{\varBodyFrameName}}}   
\newcommand{\varMagDipMomentWorld}{\ensuremath{\vect{\tilde{m}}}} 
\newcommand{\varMagDipStrength}{\ensuremath{|\vect{\tilde{m}}|}} 
\newcommand{\varR}{\ensuremath{\mat{R}}}            
\newcommand{\varPos}{\ensuremath{\vect{p}}}              
\newcommand{\varVel}{\ensuremath{\vect{v}}}              
\newcommand{\varAngvelBf}{\ensuremath{\framevect{\omega}{\varBodyFrameName}}} 
\newcommand{\varTorqueSym}{\ensuremath{\tau}} 
\newcommand{\varForceSym}{\ensuremath{f}}            
\newcommand{\varTorqueBf}{\ensuremath{\framevect{\varTorqueSym}{\varBodyFrameName}}}  
\newcommand{\varTorqueBfxy}{\ensuremath{\framevectlabelled{\varTorqueSym}{\varBodyFrameName}{xy}}} 
\newcommand{\varForce}{\ensuremath{\vect{\varForceSym}}}            
\newcommand{\varFieldWorld}{\ensuremath{\vect{b}}}            
\newcommand{\varFieldBf}{\ensuremath{\framevect{b}{\varBodyFrameName}}}
\newcommand{\varGradientWorld}{\ensuremath{\vect{g}}}         
\newcommand{\varMagMat}{\ensuremath{\specialmat{M}}}         
\newcommand{\varMagMatForceInertial}{\ensuremath{\varMagMat_g}} 
\newcommand{\varMagMatTorqueLocal}{\ensuremath{\varMagMat_b}} 
\newcommand{\varMagMatTorqueXYLocal}{\ensuremath{\varMagMat_{b \sim z}}} 
\newcommand{\varActMat}{\ensuremath{\specialmat{A}}}            
\newcommand{\varCurrents}{\ensuremath{\vect{i}}}           
\newcommand{\varAllocationMat}{\ensuremath{\mat{\Lambda}}} 
\newcommand{\varAllocationMatReduced}{\ensuremath{\mat{\bar{\Lambda}}}} 
\newcommand{\varMagMatReduced}{\ensuremath{\varMagMat_{\sim z}}}
\newcommand{\varReducedAttitude}{\ensuremath{\vect{\Gamma}}} 
\newcommand{\varBfPointingVector}{\ensuremath{\framevect{\hat{n}}{\varBodyFrameName}}} 
\newcommand{\varLevMass}{\ensuremath{m}}           

\newcommand{\varLevInertiaXX}{\ensuremath{I_{xx}}} 
\newcommand{\varLevInertiaYY}{\ensuremath{I_{yy}}} 
\newcommand{\varLevInertiaZZ}{\ensuremath{I_{zz}}} 

\newcommand{\varFullWrenchVector}{\ensuremath{
  \begin{bmatrix}
    \varTorqueBf \\
    \varForce
  \end{bmatrix}
}}

\newcommand{\varWrenchVector}{\ensuremath{
  \begin{bmatrix}
    \varTorqueBf_{xy} \\
    \varForce \\
  \end{bmatrix}
}}

\newcommand{\paramRemanence}{\ensuremath{b_r}} 
\newcommand{\constPermeabilityFreeSpace}{\ensuremath{\mu_0}} 

\newcommand{\zeromat}[1]{\ensuremath{\mat{0}_{#1}}} 

\newcommand{\varEz}{\ensuremath{\vect{e}_z}}

\newcommand{\varLQRGainSym}{\ensuremath{K}}
\newcommand{\varxLQRGain}{\ensuremath{\matsubscript{\varLQRGainSym}{x}}} 
\newcommand{\varyLQRGain}{\ensuremath{\matsubscript{\varLQRGainSym}{y}}} 
\newcommand{\varzLQRGain}{\ensuremath{\matsubscript{\varLQRGainSym}{z}}} 
\newcommand{\varxIGain}{\ensuremath{k_{\mathrm{I}x}}}
\newcommand{\varyIGain}{\ensuremath{k_{\mathrm{I}y}}}
\newcommand{\varzIGain}{\ensuremath{k_{\mathrm{I}z}}}
\newcommand{\varRADerivativeMat}{\ensuremath{\matsubscript{K}{d}}} 
\newcommand{\varRAPropotionalTerm}{\ensuremath{k_p}} 
\newcommand{\varRAIntegralTerm}{\ensuremath{k_{\mathrm{I}}}} 

\newcommand{\varSamplingTime}{\ensuremath{T_s}} 
\newcommand{\varECBCornerFrequency}{\ensuremath{f_{\mathrm{elec}}}} 

\newcommand{\controlsetpoint}[1]{\ensuremath{#1_{\scriptscriptstyle \mathrm{SP}}}} 
\newcommand{\elemcontrolsetpoint}[2]{\ensuremath{#1_{#2,\scriptscriptstyle \mathrm{SP}}}} 
\newcommand{\sptPosX}{\ensuremath{\controlsetpoint{x}}} 
\newcommand{\sptStateLinX}{\ensuremath{\elemcontrolsetpoint{\vect{s}}{x}}} 
\newcommand{\sptVelX}{\ensuremath{\elemcontrolsetpoint{v}{x}}} 
\newcommand{\sptReducedAttitude}{\ensuremath{\controlsetpoint{\varReducedAttitude}}} 
\newcommand{\sptPosition}{\ensuremath{\controlsetpoint{\varPos}}} 

\newcommand{\sptCurrents}{\ensuremath{\controlsetpoint{\varCurrents}}} 

\newcommand{\statePosX}{\ensuremath{x}} 
\newcommand{\symbStateLinX}{\ensuremath{\vect{s}_x}} 
\newcommand{\stateVelX}{\ensuremath{v_x}} 
\newcommand{\stateRoll}{\ensuremath{\phi}} 
\newcommand{\statePitch}{\ensuremath{\theta}} 
\newcommand{\stateYaw}{\ensuremath{\psi}} 
\newcommand{\stateAngvelBfX}{\ensuremath{\framevectcomp{\omega}{\varBodyFrameName}{x}}} 
\newcommand{\stateAngvelBfY}{\ensuremath{\framevectcomp{\omega}{\varBodyFrameName}{y}}} 
\newcommand{\stateAngvelBfZ}{\ensuremath{\framevectcomp{\omega}{\varBodyFrameName}{z}}} 
\newcommand{\stateAngvelBfXDot}{\ensuremath{\framevectcomp{\tderiv{\omega}}{\varBodyFrameName}{x}}} 
\newcommand{\stateAngvelBfYDot}{\ensuremath{\framevectcomp{\tderiv{\omega}}{\varBodyFrameName}{y}}} 
\newcommand{\stateAngvelBfZDot}{\ensuremath{\framevectcomp{\tderiv{\omega}}{\varBodyFrameName}{z}}} 
\newcommand{\stateAngvelBfReduced}{\ensuremath{\framevect{\bar{\omega}}{\varBodyFrameName}}} 
\newcommand{\stateRAError}{\ensuremath{\vect{e_r}}} 

\newcommand{\cinputForceX}{\ensuremath{\varForceSym_x}} 
\newcommand{\cinputTorqueX}{\ensuremath{\framevectcomp{\varTorqueSym}{\varBodyFrameName}{x}}} 
\newcommand{\cinputTorqueY}{\ensuremath{\framevectcomp{\varTorqueSym}{\varBodyFrameName}{y}}} 
\newcommand{\cinputTorqueZ}{\ensuremath{\framevectcomp{\varTorqueSym}{\varBodyFrameName}{z}}} 
\newcommand{\cvarTorqueBfxyIntegral}{\ensuremath{\framevectlabelled{\varTorqueSym}{\varBodyFrameName}{xy,I}}} 
\newcommand{\cvarForceIntegral}{\ensuremath{\varForceSym}_I} 

\newcommand{\symbNorthPole}{\ensuremath{\texttt{N}}} 
\newcommand{\symbSouthPole}{\ensuremath{\texttt{S}}} 

\newcommand{\setSOthree}{\ensuremath{\mathrm{\mathsf{SO}}(3)}} 
\newcommand{\setReal}[1]{\ensuremath{\mathbb{R}^{#1}}} 
\newcommand{\setSphere}{\ensuremath{\mathbb{S}^2}} 

\newcommand{\octomag}{OctoMag} 
\newcommand{\textsupregistered}{\textsuperscript{\sffamily\textregistered}} 

\newcommand{\pyname}[1]{\lstinline|#1|}
\tikzstyle{block} = [draw, rectangle, 
    minimum height=1.5em, minimum width=3em]
\tikzstyle{sum} = [draw, circle, node distance=1cm]
\tikzstyle{input} = [coordinate]
\tikzstyle{output} = [coordinate]
\tikzstyle{pinstyle} = [pin edge={to-,thin,black}]
\tikzstyle{connectioncircle} = [draw, circle, fill=black, radius=0.025cm]

\NewDocumentCommand {\getnodedimen} {O{\nodewidth} O{\nodeheight} m} {
  \begin{pgfinterruptboundingbox}
  \begin{scope}[local bounding box=bb@temp]
    \node[inner sep=0pt, fit=(#3)] {};
  \end{scope}
  \path ($(bb@temp.north east)-(bb@temp.south west)$);
  \end{pgfinterruptboundingbox}
  \pgfgetlastxy{#1}{#2}
}

\title{
Remote Magnetic Levitation Using Reduced Attitude Control and Parametric Field Models
}

\author{Neelaksh Singh$^\dagger$, Jasan Zughaibi$^\dagger$, Denis von Arx, Bradley J. Nelson, and Michael Muehlebach
\thanks{Manuscript received: December, 17, 2025; Revised April, 1, 2026; Accepted April, 30, 2026.}
\thanks{This paper was recommended for publication by Editor Jessica Burgner-Kahrs upon evaluation of the Associate Editor and Reviewers’ comments.
This work was supported in part by the Swiss National Science Foundation under Grant IZLCZ0\_206033, in part by the
Max Planck ETH Center for Learning Systems, and in part by the German Research Foundation.} %
\thanks{$^\dagger$These authors contributed equally to this work.}
\thanks{Corresponding author: Jasan Zughaibi}
\thanks{Neelaksh Singh, Jasan Zughaibi, Denis von Arx, and Bradley J. Nelson are with the Multi-Scale Robotics Lab, ETH Z\"urich, 8092 Z\"urich, Switzerland (e-mail: sneelaksh17@gmail.com; zjasan@ethz.ch; dvarx@ethz.ch; bnelson@ethz.ch).}%
\thanks{Michael Muehlebach is with the Learning and Dynamical Systems Group, Max Planck Institute for Intelligent Systems, 72076 T\"ubingen, Germany (email: michael.muehlebach@tuebingen.mpg.de).}%
\thanks{Digital Object Identifier (DOI): see top of this page.}
}

\usepackage{cite}

\usepackage{bm}

\begin{document}

\maketitle
\markboth{IEEE Robotics and Automation Letters. Preprint Version. Accepted May, 2026}{Singh \MakeLowercase{\textit{et al.}}: Remote Magnetic Levitation Using Reduced Attitude Control and Parametric Field Models}


\begin{abstract}

Electromagnetic navigation systems (eMNS) are increasingly used in minimally invasive procedures such as endovascular interventions and targeted drug delivery due to their ability to generate fast and precise magnetic fields.  
In this paper, we utilize the OctoMag and a custom 13-coil eMNS to achieve remote levitation and control of multiple rigid bodies across large air gaps, showcasing the dynamic capabilities of such systems.
A compact parametric analytical model maps coil currents to the forces and torques acting on the levitating object, eliminating the need for computationally expensive simulations or lookup tables and establishing a levitator- and platform-agnostic control framework.
Translational motion is stabilized using linear quadratic regulators.
A nonlinear time-invariant controller is used to regulate the reduced attitude accounting for the inherent uncontrollability of rotations about the dipole axis and stabilizing the full five degrees of freedom controllable pose subspace. 
We analyze key design limitations and evaluate the approach through trajectory tracking experiments across different objects and actuation platforms. Notably, our proposed controller demonstrates superiority over an equivalent baseline PID formulation, reliably tracking large spatial angles up to 65$^\circ$.
This work demonstrates the dynamic capabilities and potential of feedback control in electromagnetic navigation, which is likely to open up new medical applications.

\end{abstract}


\section*{Supplementary Material}
A video of our work, with additional experiments not included in this paper, can be found at \papervidurl{}.
The code accompanying this paper is available at \papercodeurl{}.

\section{Introduction}\label{sec:introduction}

\IEEEPARstart{E}{lectromagnetic} navigation systems (eMNS) are emerging as a promising technology in medical robotics, offering precise and responsive control over magnetic tools with applications ranging from endovascular interventions to targeted drug delivery \cite{dabbaghBiomedicalApplicationsMagnetic2022,fabian2025science,zhao2022tele, alex2025tele}. 
Magnetic fields are generated either by moving permanent magnets \cite{valdastri2020colon} or by driving currents through electromagnets in an eMNS \cite{gervasoni2024navion}. 
The former approach typically achieves higher field strengths and gradients, but the dynamic performance is limited by the inertia of the moving parts \cite{yangMagneticActuationSystems2020}. 
In contrast, eMNS typically have a significantly higher actuation bandwidth by rapidly modulating coil currents, making them more suitable for tasks requiring fast, closed-loop control.

The literature on clinical electromagnetic navigation has largely focused on quasi-static modeling and open-loop control \cite{cavusoglu2014MIR_quasistatic, edelmann2017magCtrlCont, fieldalign2006}.
Recent research, however, argues that exploiting the actuation bandwidth through feedback control significantly benefits procedures involving physiological motion, such as cardiac ablation \cite{zughaibiDynamicElectromagneticNavigation2025}. 
In addition, feedback control with real-time state information enables energy-efficient field allocation, increasing the workspace by up to an order of magnitude \cite{zughaibi2025expanding}.

\begin{figure}[!t]
    \includegraphics[width=\linewidth]{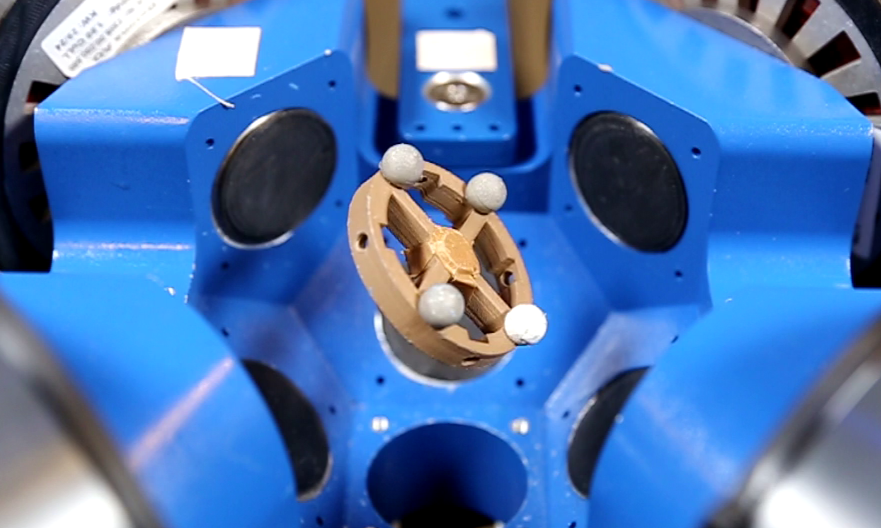}
\caption{
A freely levitating object (Object I) in the \octomag{} eMNS. 
The levitator's main body (bronze colored) was fabricated in a 3D printer using the Colorfabb BronzeFill\textsupregistered{} filament, whose high density leads to high inertia at smaller dimensions to achieve more movement space in the \octomag{}'s workspace ($\approx\varnothing$\qty{14}{\centi\meter}). A motion capture system detects three reflective markers attached to the levitator for pose estimation. The fourth marker, colored white, is non-reflective and attached solely to achieve a symmetric weight distribution.
To demonstrate that our control framework is agnostic to both the levitating object and the eMNS platform, we designed three distinct levitators. Object I (shown here) has a mass of \qty{32.4}{\gram}, principal moments of inertia $I_{xx}\approx I_{yy} = 5.9 \times 10^{-6}\,\qty{}{\kilo\gram\meter\squared}$, and a dipole made of two N52 NdFeB disc permanent magnets ($\varnothing \qty{10}{\milli\meter} \times \qty{5}{\milli\meter}$ each) symmetrically attached to its center. We also successfully levitated Object II ($m = \qty{39.4}{\gram}$, $I_{xx} \approx I_{yy} = 2.0 \times 10^{-5}\,\qty{}{\kilo\gram\meter\squared}$, identical N52 magnets) and, using a custom 13-coil eMNS, a much larger Object III ($m = \qty{209}{\gram}$, $I_{xx} \approx I_{yy} = 1.0 \times 10^{-4}\,\qty{}{\kilo\gram\meter\squared}$, containing two N45 magnets, $\varnothing \qty{30}{\milli\meter} \times \qty{10}{\milli\meter}$), as shown in the supplementary video. 
}
    \label{fig:octomag_with_steady_levitator}
\end{figure}

In this work, we further push the dynamic limits of eMNS by utilizing high-bandwidth feedback to achieve remote magnetic levitation of centimeter-scale rigid bodies with the \octomag{} eMNS \cite{kummerOctoMagElectromagneticSystem2010a}, as illustrated in \Cref{fig:octomag_with_steady_levitator}. 
Although the \octomag{} is a research-grade platform typically associated with microrobotics and ophthalmic procedures \cite{ullrichMobilityExperimentsMicrorobots2013,zhouMagneticallyDrivenMicro2021}, we demonstrate its capacity for high-bandwidth control at a macroscopic scale. To evaluate the scalability and platform-agnostic nature of our framework, we additionally demonstrate levitation on a larger 13-coil system featuring a \qty{35}{\centi\meter} workspace diameter.
Establishing high-performance remote control of magnetic objects could pave the way for new medical applications since contactless manipulation inherently reduces friction, mechanical wear, backlash, and vibrations; characteristics that are undesirable for medical applications.
For example, precise in-vivo levitation could enable diagnostic procedures such as detecting hidden tumors with sensorized ingestible capsules \cite{chenMagneticallyActuatedCapsule2022}. It could also improve colonoscopy, where magnetic actuation is already used to counteract gravity and lift a soft-tethered capsule to reduce wall contact \cite{valdastri_levitation2019}.
As clinical translation of our proposed framework relies on real-time state feedback, we emphasize the importance of high-frequency in-vivo state estimation techniques\cite{vonarxSimultaneousLocalizationActuation2024a,non_sync_cavaliereInductiveSensorDesign2020}.

To meet the listed requirements, as a proof of concept, we design a control and modeling framework to levitate rigid bodies.
We first fit an analytical model relating electrical coil currents to the magnetic field within the workspace using calibration data. The resulting field is mapped to forces and torques using an analytical point-dipole representation.
These field models are commonly used by the eMNS community as the main calibration tool \cite{charreyronModelingElectromagneticNavigation2021}; however, to the best of our knowledge, this is the first time they are utilized to achieve macroscopic levitation of a permanent magnetic dipole in a physical experimental setup.

We deploy a nonlinear, time-invariant feedback controller that guarantees stability over all controllable orientations \cite{rajStructurePreservingReduced2021,muellerStabilityControlQuadrocopter2014}. This becomes particularly important in challenging in-vivo settings (e.g. magnetic capsule in intestine) where practical operation demands control across the levitator's full range of motion. However, for a symmetric rigid body with a single embedded magnetic dipole, rotations about the dipole axis are inherently uncontrollable \cite{berkelmanMagnetLevitationTrajectory2009}. Hence, we deploy a reduced attitude representation \cite{chaturvediRigidBodyAttitudeControl2011}, that captures only the controllable attitude subspace. To evaluate the structural advantages of our approach, we benchmark the proposed controller against an equivalent decoupled proportional-integral-derivative (PID) baseline, demonstrating its superior stability and robustness during large-angle tracking and disturbance rejection.

In conclusion, this work makes three main contributions: 
First, we demonstrate, for the first time, remote magnetic levitation of a centimeter-scale rigid body in air using a clinical eMNS platform and extended working distances. 
Second, we introduce a modeling strategy that is fully agnostic to both the levitator and the electromagnetic platform. Enabled by an efficient calibration process, we demonstrate the straightforward reuse of this strategy by levitating three different objects across two separate eMNS setups.
Third, we achieve stabilization and trajectory tracking over the full five degrees-of-freedom (DoF) controllable pose subspace. We evaluate our approach through trajectory tracking experiments and discuss important design considerations of the levitation system.

\subsection{Related Work}\label{subsec:related_work__sec_introduction}
Magnetic levitation has seen a surge of research activity in recent years. Many contributions have been driven by its use in precision motion systems \cite{zhangSixAxisMagneticLevitation2007,zhouMagneticLevitationTechnology2022}, maglev transport systems \cite{kimDesignControlLevitation2017, hanMagneticLevitation2016}, magnetic suspension \cite{sawadaNAL60cmMagnetic2004}, and contactless bearings \cite{huangMagneticBearingStructure2024}.
However, remote magnetic levitation in air for biomedical applications has received little attention as most works focus on fluid-suspended magnetic bodies \cite{dabbaghBiomedicalApplicationsMagnetic2022}. 
Only a few recent works explore remote magnetic levitation targeted towards biomedical settings \cite{non_sync_zhengLowpowerMagneticLevitation2025}.
Therefore, we mainly review advances in remote macroscopic magnetic levitation in air from non-medical domains that address the key challenges of large motion ranges and high-bandwidth control, which are important criteria for prospective clinical applications.

In the context of agile and high-range 5-6 DoF permanent magnet levitation, many works employ planar eMNS architectures that can be tiled to extend the horizontal workspace \cite{lu6DDirectdriveTechnology2012a, dyckMagneticallyLevitatedRotary2017, berkelmanMagnetLevitationTrajectory2009,non_sync_zhuDesignModelingSixDegreeofFreedom2017}. 
In \cite{berkelmanMagnetLevitationTrajectory2009}, a magnet is levitated using 10 cylindrical coils arranged on a plane. 
Subsequent works \cite{berkelmanMultipleMagnetIndependent2023, miyasakaMagneticLevitationUnlimited2014a} build on this platform to demonstrate 6-DoF control of an object with two independent dipoles, simultaneous levitation of two such levitators, and omnidirectional control of a levitator with six spherically arranged permanent magnets. 
Similarly, \cite{wangMagFloorUniversalMagnetic2024} introduces a planar eMNS with square coils that simultaneously levitates a 2D Halbach array and an object with three magnets.
The recently commercialized X-Planar\textregistered{} system by Beckhoff Automation allows fast 6-DoF control of platforms with magnet arrays with an unrestricted horizontal range \cite{bentfeldPlanarDriveSystem2025}.
In all these works, pose control is achieved through decoupled PID controllers which leads to a restricted range of stable attitudes.
While the resulting motion range is sufficient for many industrial applications often dealing with controlled environments, clinical applications will require stabilization over a larger attitude range.
Complex magnet configurations have been used to increase the range of stable attitudes in \cite{miyasakaMagneticLevitationUnlimited2014a} with complex digital estimation and control.
We instead rely on a simple single magnet levitator design and use a control scheme that leads to Lyapunov stability of almost the entire controllable attitude range.

A further distinction between previous works and our approach lies in the modeling paradigm. 
Existing high-range levitation systems predominantly identify or derive a direct map from coil currents to forces and torques on a specific levitator, often using force-torque sensors, levitator specific data collection rigs, and finite element analysis  \cite{berkelmanMagnetLevitationTrajectory2009,non_sync_xuRealTimeDataDrivenForce2022,non_sync_wangDeepLearningBasedWrench2024a,berkelmanMultipleMagnetIndependent2023}. 
This leads to models that must be re-identified whenever the levitator's magnet configuration changes. 
In contrast, the field-centric modeling strategy in this work decouples eMNS calibration from the specific magnet configuration, and, as long as independent magnet volumes remain small compared to the coil dimensions, new levitator designs can be accommodated by updating only their magnetic dipole parameters rather than re-identifying a full current to force-torque map.

\subsection{Outline}\label{subsec:outline__sec_introduction}
The rest of this paper is organized as follows: 
\Cref{sec:experimental_setup_and_field_model} describes the experimental setup and gives a brief overview of the field model which relates coil currents to forces and torques on the levitating object.
The dynamics of the levitator and control design are presented in \cref{sec:dynamics_and_control}. 
Key performance considerations of the levitation system and the results from trajectory tracking experiments are discussed in \cref{sec:results_and_discussion}.
Finally, a conclusion is drawn in \cref{sec:conclusion}.
\begin{figure*}[!t]
    \centering
    \subfloat[]{\label{fig:octomag_levitation_coordinate_systems}%
        \centering
        \scalebox{0.9}{
            \begin{tikzpicture}[yscale=-1]

                \clip (2,0) rectangle (10,6);

                \node[anchor=north west, inner sep=0] (img) at (0,0) {
                    \includegraphics[width=12cm]{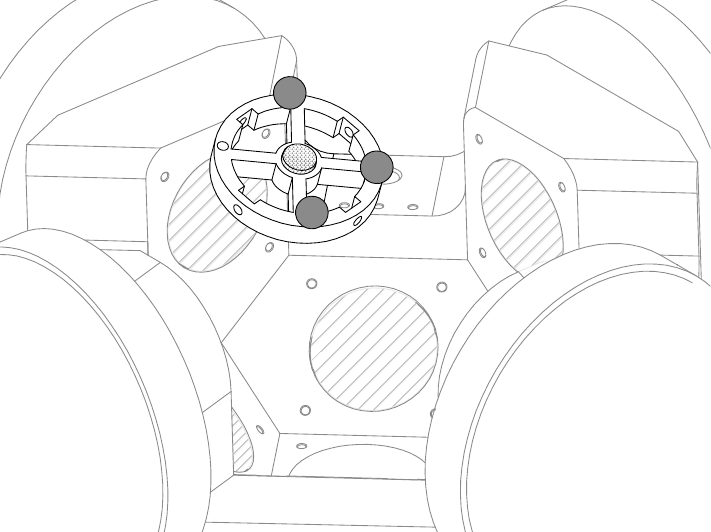}
                };


                \coordinate (V) at (6,5,0);
                \coordinate (C1_center) at (8.735, 3.641, 0);
                \coordinate (C2_center) at (3.688, 3.617, 0);

                \coordinate (C1_core_label_point) at ($(C1_center) + (-0.25, 0.0, 0.0)$);
                \coordinate (core_label_point) at ($(C1_center) + (0.5, -1.6, 0.0)$);
                \draw[thick] (C1_core_label_point) -- (core_label_point) node[anchor=south, align=center, font=\footnotesize] {Soft iron \\ core};

                \coordinate (coil1_label_point) at (9.197, 0.599, 0);
                \coordinate (coil3_label_point) at (3.701, 0.506, 0);
                \coordinate (coil_label_point) at (6.473, 0.292, 0);
                \node[font=\footnotesize, align=center, name=coil_label_node] at (coil_label_point) {Solenoidal coils};
                \draw[thick] (coil1_label_point) -- (coil_label_node);
                \draw[thick] (coil3_label_point) -- (coil_label_node);
                
                \draw[dashed, gray] (V) -- (C1_center);
                \draw[dashed, gray] (V) -- (C2_center);
                \draw pic [draw, gray, angle radius=2mm] {right angle=C1_center--V--C2_center};

                \draw[dashed, gray] (V) -- (6.41,3.414,0); 
                \draw[thick,->, BrickRed] (V) -- (6.1759038,4.31537449,0) node[below right=-2]{$\vect{x}$};
                \draw[thick,->, Green] (V) -- (5,5,0) node[below left=-3]{$\vect{y}$};
                \draw[thick,->, blue] (V) -- (6,4,0) node[above right=-3.6]{$\vect{z}$};
                \draw[fill] (V) circle (0.05) node[below right=-1]{$\varWorldFrame$};

                \coordinate (B) at (4.978, 2.831, 0);
                \draw[dashed, gray] (4.496, 3.173, 0) -- (4.022, 3.56, 0);
                \draw[dashed, gray] (4.476, 2.404, 0) -- (4.093, 2.050, 0);
                \draw[thick, ->, BrickRed] (B) -- (4.476, 2.404, 0) node[left=-2]{$\vect{x}$};
                \draw[thick, ->, Green] (B) -- (4.496, 3.173, 0) node[below right=-2]{$\vect{y}$};
                \coordinate (z_bf_end) at (5.110, 2.389, 0);
                \draw[thick, ->, blue] (B) -- (z_bf_end);
                \node[blue] at ($(z_bf_end) + (-0.05, -0.1)$) {$\vect{z}$};
                \draw[fill] (B) circle (0.05) node[above right=25, name=bframe_label, inner sep=1pt]{$\varBodyFrame$};
                \draw[thick] (B) -- (bframe_label.south);

                \draw[thick, <-, cyan, dashed] (B) .. controls +(120:1cm) and +(-120:1cm) .. (V)
                node[pos=0.71, inner sep=2pt, fill=white]{\footnotesize{$\varR, \varPos$}};

        \end{tikzpicture}
        }
    }%
    \hspace{2.0cm}
    \subfloat[]{\label{fig:levitator_top_view_and_cross_section}%
        \centering
        \scalebox{0.7}{
            \begin{tikzpicture}[yscale=-1]


                \node[anchor=north west, inner sep=0] (img) at (0,0) {
                    \includegraphics{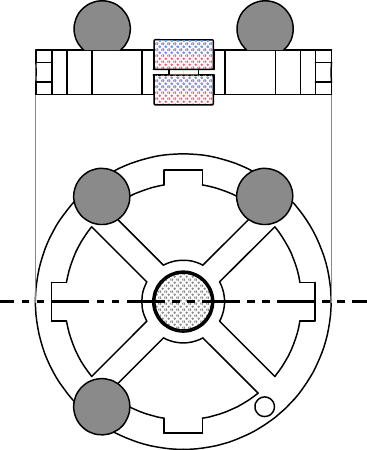}
                };

                \coordinate (B) at (3.1002, 1.2190);
                \draw[thick, NavyBlue, ->] (B) -- ($(B) + (0, 1)$) node[right=-2.0] {$\varMagDipMoment$};
                \draw[blue, thick, ->] (B) -- ($(B) + (0,-1)$) node[above=0]{$\vect{z}$};
                \draw[Green, thick, ->] (B) -- ($(B) + (1, 0)$) node[above=-1]{$\vect{y}$};
                \node at ($(B) + (0.5, -0.8)$) {$\varBodyFrame$};
                \draw[fill, Red] (B) circle (0.05);
                \draw[BrickRed, thick] (B) circle (0.2) node[below right=1]{$\vect{x}$};

                \coordinate (B_top) at (3.100, 5.1012);
                \draw[BrickRed, thick, ->] (B_top) -- ($(B_top) + (0,1)$) node[below]{$\vect{x}$};
                \draw[Green, thick, ->] (B_top) -- ($(B_top) + (1, 0)$) node[above=-1.0]{$\vect{y}$};
                \draw[fill, blue] (B_top) circle (0.05);
                \draw[blue, thick] (B_top) circle (0.2) node[above=3.0]{$\vect{z}$};

                \coordinate (BMC) at ($(B) + (-0.2, 0.3)$);
                \coordinate (M_label_pt) at ($(BMC) + (-0.7, 0.3)$);
                \draw[thick] ($(BMC) + (-0.0, -0.5)$) -- (M_label_pt);
                \draw[thick] (BMC) -- (M_label_pt) node[below left=-3] {Magnets};

                \coordinate (VMC) at ($(B) + (1.386, -0.844)$);
                \coordinate (VMC2) at (4.4821, 3.3073);
                \coordinate (V_label_pt) at ($(VMC) + (0.8, 1.5)$);
                \node[fill=white] (v_label) at (V_label_pt) [below = -1] {Reflective markers};
                \draw[thick] (VMC) -- (v_label.north);
                \draw[thick] (VMC2) -- (v_label.south);

                \node[Blue, thick, above=-1] at (2.6087, 0.6750) {$\symbSouthPole$};
                \node[Red, thick, below=-1] at (2.6087, 1.7670) {$\symbNorthPole$};

        \end{tikzpicture}
        }
    }%
    \caption{
    Geometric notations and key components of the eMNS and the levitator. 
    (a) The levitator's body frame $\varBodyFrame$ and the inertial frame $\varWorldFrame$.
    The \octomag{} consists of eight solenoidal coils (approx. $\varnothing \qty{103}{\milli\meter} \times \qty{170}{\milli\meter}$), each with a soft iron core, arranged in a spherical combination.
    The origin of \varWorldFrame{} is located at the center of this sphere.
    Full pose of the levitator is provided as the rotation matrix $\varR \in \setSOthree$ from $\varBodyFrame$ to $\varWorldFrame$, and the position vector $\varPos \in \setReal{3}$ of $\varBodyFrame$ expressed in $\varWorldFrame$. 
    (b) The top and cross-section views of the levitator illustrating the placement of permanent magnets.
    Three reflective markers from \cref{fig:octomag_with_steady_levitator} are shown in gray, while the non-reflective marker is not shown. 
    The origin of $\varBodyFrame$ is approximately at the center of mass of the levitator. 
    The two axially magnetized disc magnets used in the levitator are illustrated with a dot pattern. 
    The net magnetic dipole moment of the assumed ideal point dipole in \varBodyFrame{} is $\varMagDipMoment = -\varMagDipStrength \framevectlabelled{e}{\varBodyFrameName}{z}$, since dipole moment is oriented from the magnet stack's overall south-pole (\symbSouthPole) to the north-pole (\symbNorthPole) by convention.}
    \label{fig:levitator_object_octomag_diagrams}
\end{figure*} 
\section{Experimental Setup and Field Model}\label{sec:experimental_setup_and_field_model}

This section describes the hardware setup used to develop and test the levitation pipeline, followed by an overview of the mathematical models governing the influence of coil currents on forces and torques applied to a permanent magnetic dipole.

\subsection{Notation}
\label{subsec:notation__sec_experimental_setup_and_results}
Vectors are represented by bold lowercase letters, e.g. $\vect{x}$, and matrices are represented by bold uppercase or calligraphic letters, e.g. $\mat{X}$ or $\specialmat{X}$.
We use two coordinate frames: the body-fixed frame referred to as \varBodyFrame{}, and the inertial world frame referred to as \varWorldFrame{}.
Any vector $\vect{x}$ resolved in \varBodyFrame{} is denoted by $\framevect{x}{B}$, while all vectors without a frame superscript are expressed in \varWorldFrame{}.
The set of all normalized vectors in \setReal{3} is the unit sphere represented by \setSphere{}. 
The special orthogonal group, represented by \setSOthree{}, is the set of rigid rotations in \setReal{3}.
We denote the standard unit vector along the $z$ axis of any frame as $\varEz = \transpose{\begin{bmatrix} 0 & 0 & 1 \end{bmatrix}}$.
Forces and torques are collectively referred to as a wrench, following standard convention in robotics.

\subsection{Experimental Setup}
\label{subsec:experimental_setup__sec_experimental_setup_and_results}
We evaluate our framework using two distinct eMNS platforms to demonstrate its hardware-agnostic capabilities: the \octomag{} eMNS, which comprises eight symmetrically arranged coils with a \qty{14}{\centi\meter} workspace, and a larger custom eMNS featuring 13 coils and a \qty{35}{\centi\meter} workspace diameter. Both systems share an identical coil architecture, with each coil consisting of 1400 turns of laminated copper wire wound around a soft iron core. To achieve sufficient current controller bandwidth, a challenging task given the large coil inductance, we utilize the same custom-built drivers for both platforms.

These drivers regulate coil currents using pulse-width modulated (PWM) voltages at a frequency of \qty{20}{\kilo\hertz} and an amplitude of \qty{100}{\volt}. The effective bandwidth is characterized by the corner frequency (\varECBCornerFrequency{}), defined as the -\qty{3}{\decibel} point of the closed-loop response. This frequency determines the speed of current regulation and is a critical parameter in cascaded control architectures, where maximizing inner-loop bandwidth is generally desirable. Using the system identification method in \cite{zughaibiDynamicElectromagneticNavigation2025}, we measured a corner frequency of $\varECBCornerFrequency=\qty{26.4}{\hertz}$ at a current amplitude of \qty{5}{\ampere}. This bandwidth is significantly lower than those reported for smaller coils in related work \cite{berkelmanMagneticLevitationLarge2013,zhangMagTableTabletopSystem2021}, underscoring the growing difficulty of achieving stable magnetic levitation as coil dimensions increase. Crucially, levitating lower-inertia objects necessitates higher-bandwidth coil drivers, as their shorter mechanical time constants require faster stabilizing control responses. Consequently, clinical translation for applications such as swallowable capsules will rely on either advancing these driver technologies or utilizing coils with reduced inductance. To demonstrate that our approach is agnostic to the levitating object, we designed three distinct 3D-printed levitators (Objects I--III) spanning diverse inertia ranges. Object I is depicted in \Cref{fig:octomag_with_steady_levitator}. Detailed physical parameters for all three objects are provided in the caption of \Cref{fig:octomag_with_steady_levitator}, while \Cref{fig:levitator_object_octomag_diagrams} provides an overview of relevant coordinate frames and key mechanical components.

A Vicon\textregistered{} motion-capture system, running Tracker\textregistered{} software and equipped with five T-10S cameras operating at \qty{1}{\kilo\hertz}, tracks the levitator's pose in real time by localizing three retro-reflective markers mounted on the levitator (see \Cref{fig:levitator_object_octomag_diagrams}).

\subsection{Magnetic Field Model}\label{subsec:magnetic_field_model__sec_modelling}
The soft iron cores used in the coils of both platforms exhibit negligible hysteresis effect \cite{kummerOctoMagElectromagneticSystem2010a}.
We limit the operating current in each coil to $\pm$ \qty{4}{\ampere} primarily to protect the hardware electronics; consequently, the cores remain far from magnetic saturation.
Therefore, the magnetic fields $\varFieldWorld \in \setReal{3}$ and gradients $\varGradientWorld =  \transpose{\begin{bmatrix} \partialderivative{b_x}{x} & \partialderivative{b_x}{y} & \partialderivative{b_x}{z} & \partialderivative{b_y}{y} & \partialderivative{b_y}{z} \end{bmatrix}} \in \setReal{5}$ at any position $\varPos \in \setReal{3}$ can be assumed to vary linearly with coil currents $\varCurrents \in \setReal{N}, N\in\{ 8, 13\}$ as follows
\begin{equation}
    \begin{bmatrix}
        \varFieldWorld \\
        \varGradientWorld
    \end{bmatrix} = \func{\varActMat}{\varPos} \varCurrents,\qquad \varActMat \in \setReal{8 \times N}, 
    \label{eq:currents_to_field_gradient_map}
\end{equation}
where $\varFieldWorld, \varGradientWorld$, and $\varPos$ are expressed in \varWorldFrame{} by convention \cite{petruskaMinimumBoundsNumber2015}, and $\varActMat$ is the so-called actuation matrix.
Note that there are only five independent gradient components for the magnetic field due to Maxwell's laws for quasi static fields in the absence of free currents ($\nabla \cdot \varFieldWorld = 0$ and $\nabla \times \varFieldWorld = 0$).

The actuation matrix is derived from the field model, with each column representing the field and gradient at position $\varPos$ due to a unit current in the corresponding coil.
Obtaining this model is the calibration step, which can be performed using various eMNS modelling methods \cite{charreyronModelingElectromagneticNavigation2021}, of which we use the Multipole Expansion Model (MPEM) \cite{petruskaModelBasedCalibrationMagnetic2017, bernardes2026structured}.
MPEM is a parametric model derived from the analytical solution for cylindrically symmetric systems and models each coil as the multipole expansion of a point source with its magnetic center, orientation, and strength as parameters to be estimated.
A simplified representation neglecting the higher order terms in the multipole expansion is sufficient due to the nominally large air gaps between the coil cores and the levitator in this system.
Therefore, for our purpose using only the dipole term is sufficient, and we assume that any modelling inaccuracies can be compensated through the feedback controller.
This method requires only a few kilobytes of memory, can be computed offline via data-based parameter optimization, and can be queried in real-time to obtain $\func{\varActMat}{\varPos}$ at any arbitrary position.
Parameter optimization is performed using a least-squares fit to magnetic field measurements as described in \cite{petruskaModelBasedCalibrationMagnetic2017}.
Data was collected for various currents in each coil at \qty{320}{} positions in \varWorldFrame{} with a calibration cube comprising a $4\times4\times4$ grid of Melexis MLX90393\textsupregistered{} hall sensors, yielding an RMS calibration error of less than \qty{0.15}{\milli\tesla} for both eMNSs.

The levitator's magnetic volume is several orders of magnitude smaller than the dimensions of the electromagnets and the operating workspace so it can be approximated as an ideal point dipole at the magnetic volume's centroid which coincides with the levitator's center of mass as shown in \cref{fig:levitator_top_view_and_cross_section}\footnote{This assumption was supported using a magnet volume discretization analysis similar to \cite{zughaibiDynamicElectromagneticNavigation2025}. The analysis yielded an average relative RMSE discrepancy of 7.7\% across all wrench components, confirming these deviations are small enough to be handled by the feedback controller.}.
The body frame torques \varTorqueBf{} acting on the point magnetic dipole can be computed as 
$
        \varTorqueBf = \varMagDipMoment \times \varFieldBf = \skewmat{\varMagDipMoment} \transpose{\varR} \varFieldWorld = \varMagMatTorqueLocal \varFieldWorld,
$
where $\skewmat{\cdot}$ denotes the skew-symmetric matrix operator, \varMagDipMoment{} is the magnetic dipole moment, and \varR{} the rotation from \varBodyFrame{} to \varWorldFrame{}.
The expression for force is given by $\varForce = \left( \varMagDipMomentWorld \cdot \nabla \right) \varFieldWorld$ and can be simplified as \cite{abbottMagneticMethodsRobotics2020}:
\begin{equation}
    \begin{aligned}
        \varForce &= \begin{bmatrix}
            m_x &m_y &m_z & 0 & 0 \\
            0 &m_x & 0 &m_y &m_z \\
            -m_z & 0 &m_x & -m_z &m_y
        \end{bmatrix} \varGradientWorld \coloneqq \varMagMatForceInertial \varGradientWorld.
    \end{aligned}
    \label{eq:gradients_to_force_inertial_map}
\end{equation}
Hence, the overall map from the coil currents to the wrench acting on the levitating object can be expressed as
\begin{equation}
    \begin{aligned}
        \varFullWrenchVector &=
        \begin{bmatrix}
        \varMagMatTorqueLocal & \zeromat{3 \times 5} \\
        \zeromat{3 \times 3} & \varMagMatForceInertial
        \end{bmatrix}
        \begin{bmatrix}
            \varFieldWorld \\
            \varGradientWorld
        \end{bmatrix} =
        \func{\varMagMat}{\varR, \varMagDipMoment}
        \begin{bmatrix}
            \varFieldWorld \\
            \varGradientWorld
        \end{bmatrix} \\
        &= \func{\varMagMat}{\varR, \varMagDipMoment} \func{\varActMat}{\varPos} \varCurrents \coloneq \func{\varAllocationMat}{\varR, \varMagDipMoment, \varPos} \varCurrents\,,
    \end{aligned}
    \label{eq:currents_to_wrench_map}    
    \vspace{-2mm}
\end{equation}
where $\funcdef{\varMagMat}{\setSOthree \times \setReal{3}}{\setReal{6 \times 8}}$ is known as the magnetic interaction matrix.
We define the overall map from currents to wrench by $\varAllocationMat$, hereafter referred to as the allocation matrix.
The functional dependence of $\varAllocationMat, \varMagMat$, and $\varActMat$ on the orientation \varR{}, position \varPos{}, and dipole moment \varMagDipMoment{} is considered implicit in the notation and will be omitted in equations henceforth.

There is a zero row in $\skewmat{\varMagDipMoment}$ as magnetic fields cannot apply torques about the dipole axis, so $\cinputTorqueZ = 0$ for any set of coil currents.
Therefore, we can consider a reduced map with the uncontrollable torque component removed as follows
\begin{align}
    \varMagMatTorqueXYLocal &=
        \begin{bmatrix}
            0 & \varMagDipStrength & 0 \\
            -\varMagDipStrength & 0 & 0 \\
        \end{bmatrix} \transpose{\varR}
        \label{eq:magnetic_interaction_matrix_xy_torque_local} \\
    \varWrenchVector &= \varAllocationMatReduced \varCurrents \,, \hspace{2mm} \varAllocationMatReduced = \begin{bmatrix}
        \varMagMatTorqueXYLocal & \zeromat{2 \times 5} \\
        \zeromat{3 \times 3} & \varMagMatForceInertial
    \end{bmatrix} \varActMat = \varMagMatReduced \varActMat \label{eq:reduced_allocation_matrix} 
\end{align}
where $\varTorqueBfxy = \transpose{\begin{bmatrix} \cinputTorqueX & \cinputTorqueY \end{bmatrix}}$, $\varMagMatReduced$ is the reduced magnetic interaction matrix, and $\varAllocationMatReduced$ is the reduced allocation matrix. 
The dipole strength $\varMagDipStrength$ can be estimated by {$\varMagDipStrength = \paramRemanence V/\constPermeabilityFreeSpace$}, where $\paramRemanence$ is the remanence of the magnet's material, $V$ is the net magnetic volume, and $\constPermeabilityFreeSpace$ is vacuum magnetic permeability.
As the levitator's mass increases, a larger magnetic volume with higher dipole strength is required to levitate while keeping currents within limits.
This also applies when levitating close to the edges of the workspace where even small forces and torques can demand high currents \cite{boehlerWorkspaceElectromagneticNavigation2023}.
However, an increase in the net magnetic volume means the ideal point dipole assumption becomes less accurate.
\section{Dynamics and Control}\label{sec:dynamics_and_control}

For all practical purposes, the levitator's velocity stays sufficiently small so that aerodynamic drag can be neglected.
Then the dynamics of our levitator can be described as follows 
\begin{align}
    \tderiv{\varPos} &= \varVel, \qquad
    \tderiv{\varVel} = \varLevMass^{-1} \varForce, \qquad
    \tderiv{\varR} = \varR \skewmat{\varAngvelBf} \label{eq:rotation_and_linear_kinematics} \\
    \stateAngvelBfXDot &= \varLevInertiaXX^{-1}(\varLevInertiaYY - \varLevInertiaZZ) \stateAngvelBfY \stateAngvelBfZ + \varLevInertiaXX^{-1}\cinputTorqueX \label{eq:diagonal_inertia_dynamics_omega_x_dot} \\
    \stateAngvelBfYDot &= \varLevInertiaYY^{-1}(\varLevInertiaZZ - \varLevInertiaXX) \stateAngvelBfX \stateAngvelBfZ + \varLevInertiaYY^{-1}\cinputTorqueY \label{eq:diagonal_inertia_dynamics_omega_y_dot} \\
    \stateAngvelBfZDot &= \varLevInertiaZZ^{-1}(\varLevInertiaXX - \varLevInertiaYY) \stateAngvelBfX \stateAngvelBfY \label{eq:diagonal_inertia_dynamics_omega_z_dot}
\end{align}
where $\varVel \in \setReal{3}$ is the levitator's linear velocity in \varWorldFrame{}, $\varAngvelBf \in \setReal{3}$ is its angular velocity in \varBodyFrame{}, \varLevMass{} its mass, and $\varLevInertiaXX, \varLevInertiaYY, \varLevInertiaZZ$ its principal moments of inertia \cite{chaturvediRigidBodyAttitudeControl2011}.
Because our levitators are symmetric about the $x$ and $y$ axes ($\varLevInertiaXX \approx \varLevInertiaYY$), the dynamics of $\stateAngvelBfZ$ become inaccessible, preventing full attitude stabilization \cite{krishnanAttitudeStabilizationRigid1994}.
We can still stabilize a reduced version of the attitude which removes the uncontrollable rotation mode \cite{muellerStabilityControlQuadrocopter2014}.
\begin{figure*}[!t]
    \centering
    \includegraphics[clip, trim=1.9cm 21.7cm 2.0cm 1.7cm, width=\textwidth]{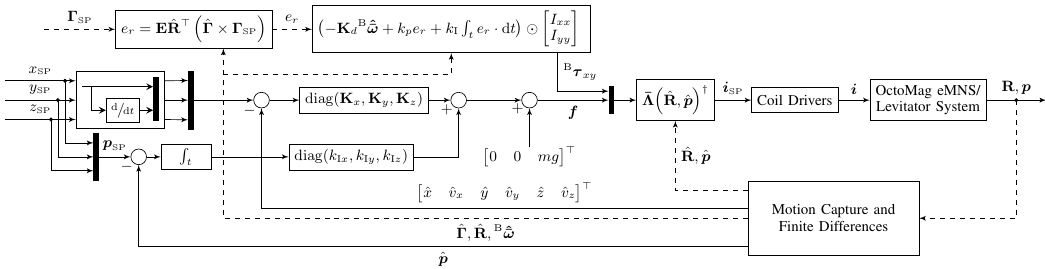}
    \caption[Block diagram of the full levitation pipeline]{Block diagram of the full levitation pipeline. 
    All state variables with a $\hat{}$ in the diagram are state estimates from motion capture feedback and finite differences.
    Solid lines between blocks represent signals with linear gains and operations while dashed lines represent flow of information through non-linear function blocks.
    The solid black vertical bars represent column vector concatenation blocks.
    }
    \label{fig:full_levitation_pipeline_block_diagram}
\end{figure*}

\subsection{Feedback Controller Design}\label{subsec:feedback_controller_design__sec_dynamics_and_control}
We use the definition of reduced attitude from \cite{rajStructurePreservingReduced2021}, which relies on choosing a fixed direction $\varBfPointingVector \in \setSphere$ in $\varBodyFrame$. This representation disregards all rotations of the levitator about the axis $\varBfPointingVector$. By choosing $\varBfPointingVector = \framevectlabelled{e}{B}{z}$, the reduced attitude $\varReducedAttitude \in \setSphere$ and its resulting kinematics are given by
\begin{align*}
    \varReducedAttitude &= \varR \varBfPointingVector, \qquad
    \tderiv{\varReducedAttitude} = \left(\varR \begin{bmatrix} \stateAngvelBfReduced \\ 0 \end{bmatrix}\right) \times \varReducedAttitude\,, \qquad \stateAngvelBfReduced = \begin{bmatrix} \stateAngvelBfX \\ \stateAngvelBfY \end{bmatrix}. \label{eq:reduced_attitude_kinematics}
\end{align*}

The reduced attitude \varReducedAttitude{} explicitly removes the inaccessible $\stateAngvelBfZ$ component from the kinematics. 
A non-linear time-invariant controller is defined to stabilize the reduced attitude to a desired value \sptReducedAttitude{} with the two available torques as follows
\begin{equation}
        \begin{bmatrix}
            \cinputTorqueX \\ \cinputTorqueY
        \end{bmatrix} = \left(-\varRADerivativeMat \stateAngvelBfReduced + \varRAPropotionalTerm \mat{E} \transpose{\varR} \left( \varReducedAttitude \times \sptReducedAttitude \right)\right) \odot \begin{bmatrix}
            \varLevInertiaXX \\ \varLevInertiaYY
        \end{bmatrix} \,, \\
    \label{eq:reduced_attitude_controller}
\end{equation}
where $\mat{E} = \begin{bmatrix} \bm{\mathrm{I}}_2 & \bm{0}_{2 \times 1}\end{bmatrix}, \varRADerivativeMat \in \setReal{2 \times 2}, \varRADerivativeMat \succ 0, \varRAPropotionalTerm > 0$, $\odot$ is the element-wise product, and $\varRADerivativeMat$ and $\varRAPropotionalTerm$ are tuning parameters. To tune these parameters, a base ratio between $\varRAPropotionalTerm$ and $\varRADerivativeMat$ is chosen to prevent under-damped oscillations, followed by a uniform scalar increase to maximize bandwidth just below the onset of noise amplification.
\Cref{eq:reduced_attitude_controller} renders the closed-loop reduced attitude dynamics  asymptotically stable about the desired equilibrium $(\sptReducedAttitude, 0)$ with a region of attraction that contains the entire state space except for the unstable antipodal equilibrium $(-\sptReducedAttitude, 0)$; see \cite{rajStructurePreservingReduced2021} for a formal analysis. Consequently, the controller can track any arbitrary setpoint $\sptReducedAttitude \in \setSphere$ for all initial conditions except $(\varReducedAttitude_0, \varAngvelBf_0) = (-\sptReducedAttitude, 0)$.
To benchmark the structural advantages of the proposed nonlinear formulation, we derive an equivalent decoupled proportional-derivative (PD) baseline using a small-angle approximation of \Cref{eq:reduced_attitude_controller}. By extracting the yaw-invariant XYZ intrinsic Euler angles $\phi$ (roll) and $\theta$ (pitch) directly from the body's normal vector, the baseline PD controller is defined as:
\begin{equation}
    \begin{bmatrix}
        \cinputTorqueX \\ \cinputTorqueY
    \end{bmatrix}_{\text{PD}} = \left(-\varRADerivativeMat \stateAngvelBfReduced + \varRAPropotionalTerm \begin{bmatrix}
        \phi_{\text{SP}} - \phi \\ \theta_{\text{SP}} - \theta
    \end{bmatrix} \right) \odot \begin{bmatrix}
        \varLevInertiaXX \\ \varLevInertiaYY
    \end{bmatrix} \,,
    \label{eq:PD_baseline}
\end{equation}
where $\phi_{\text{SP}}$ and $\theta_{\text{SP}}$ are the desired setpoint angles. By retaining the identical tuning gains $\varRADerivativeMat$ and $\varRAPropotionalTerm$, this formulation ensures the PD baseline provides mathematically equivalent control effort near the hover equilibrium.

The translational dynamics of the levitator are modeled as three decoupled double integrators; for instance $m \ddot{x} = f_x$ along the x-axis. Let $v_x \coloneqq \dot{x}$ and $\symbStateLinX = \transpose{\begin{bmatrix} \statePosX & \stateVelX \end{bmatrix}}$ denote the corresponding state vector. 
Expressing this as a first-order state-space model and applying a zero-order hold discretization with the controller's sampling time $\varSamplingTime$ yields
$
    \dfunc{\symbStateLinX}{k+1} = \mat{A}\dfunc{\symbStateLinX}{k} + \mat{B}\dfunc{\cinputForceX}{k}.
$
We use this model to design a discrete-time LQR controller to compute the force $\cinputForceX$ for regulating the state $\symbStateLinX$ to a desired setpoint $\sptStateLinX = \transpose{\begin{bmatrix} \sptPosX & \sptVelX \end{bmatrix}}$ as
$
    \dfunc{\cinputForceX}{k} = \varxLQRGain \left( \sptStateLinX - \symbStateLinX \right)\,,
$ where $\varxLQRGain \in \setReal{2}$ is the LQR gain matrix for the $x$ axis.
The same procedure applies to the $y$ and $z$ axes, and results in the LQR gain matrices $\varyLQRGain, \varzLQRGain \in \setReal{2}$.

We compute the desired coil currents $\sptCurrents \in \setReal{8}$ from the desired wrench as $ \sptCurrents = \pseudoinverse{\varAllocationMatReduced} \begin{bmatrix}
    \varTorqueBf_{xy}^\top &
    \varForce^\top 
  \end{bmatrix}^\top,$ where $\pseudoinverse{\varAllocationMatReduced}$ is the pseudoinverse of the reduced allocation matrix. Since the pseudoinverse yields the minimum 2-norm solution for the currents, and all coils possess identical ohmic resistance, this approach guarantees minimal thermal power dissipation to achieve the desired wrench.
These current setpoints are then sent to the coils' electrical drivers which use a proportional integral (PI) controller for tracking the actual currents $\varCurrents$.
Any inaccuracy in the field model, and thus $\varAllocationMatReduced$, lead to input disturbances that cause steady-state errors despite inherent integrators in the dynamics.
To address this, we design integral actions, $\cvarTorqueBfxyIntegral$ for the reduced attitude controller and $\cvarForceIntegral$ for the translational controllers, as follows:
\begin{align}
    \cvarTorqueBfxyIntegral & = \left(\varRAIntegralTerm \int_t \stateRAError \cdot \differential{t} \right) \odot \begin{bmatrix}
        \varLevInertiaXX \\ \varLevInertiaYY
    \end{bmatrix} \label{eq:reduced_attitude_controller_integral_action} \\
    \cvarForceIntegral & = \diag{\varxIGain, \varyIGain, \varzIGain} \int_t \left( \sptPosition - \varPos \right) \differential{t} + \varLevMass g \bm{e}_z
     \,,\label{eq:linear_control_integral_action_and_gravity_compensation}
\end{align}
where $\varRAIntegralTerm, \varxIGain, \varyIGain$ and $\varzIGain$ are integrator gains, $\sptPosition$ is the desired position, and $\stateRAError = \mat{E} \transpose{\varR} \left( \varReducedAttitude \times \sptReducedAttitude \right)$ is the error in reduced attitude, and $\varLevMass g \bm{e}_z$ provides gravity feedforward compensation.
Input disturbance due to model inaccuracies vary with the levitator's pose; therefore, the integral action must respond rapidly to remove steady-state offsets. For a mathematical stability analysis of the integral controller in combination with the reduced attitude controller, the reader is referred to the Online Appendix~\cite{online_appendix}. The resulting controller is shown in \cref{fig:full_levitation_pipeline_block_diagram}. An equivalent integral action is integrated into the PD baseline (see \eqref{eq:PD_baseline}) to form a full PID controller. This is achieved by substituting $\stateRAError$ with the linear Euler angle error vector $[\phi_{\text{SP}} - \phi, \theta_{\text{SP}} - \theta]^\top$.

To achieve robust levitation, the pose controller's sampling time $\varSamplingTime$ is critical. While the drivers' PWM signals cause high-frequency ripples in the coil currents and fields, these do not affect the levitator due to its inertia. Consequently, the levitator is driven by the mean coil currents. Because the digital controller acts as a zero-order hold, these mean currents remain constant between sampling instants, subjecting the levitator to momentarily static magnetic fields. According to Earnshaw's theorem, such static fields inherently exhibit negative position stiffness \cite{schweitzer2009magnetic}. This uncompensated negative stiffness causes the system state to diverge between samples, significantly reducing the actual tolerable time delay of the closed-loop system. Therefore, $\varSamplingTime$ must be kept as low as possible. We operate the pose controllers at \qty{1}{\kilo\hertz}, meaning $\varSamplingTime=\qty{1}{\milli\second}$, which is fast enough to allow for stable levitation using simple controller tuning without explicitly modeling this negative stiffness.
\begin{figure*}[!t]
    \centering
    \subfloat{%
        \includegraphics[width=0.47\linewidth, trim=5 5 5 1, clip]{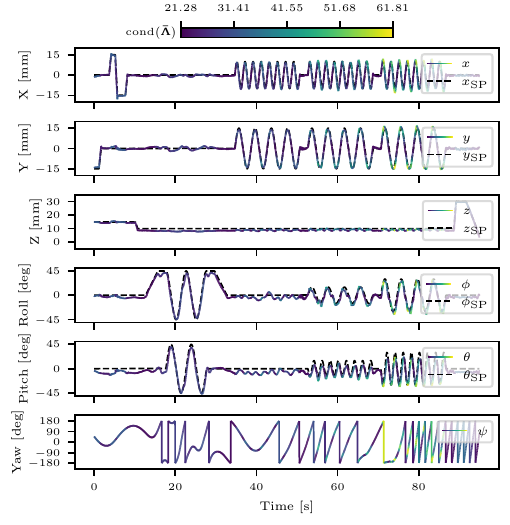}%
    }%
    \hfill
    \subfloat{%
        \includegraphics[width=0.51\linewidth, trim=5 7 1 7, clip]{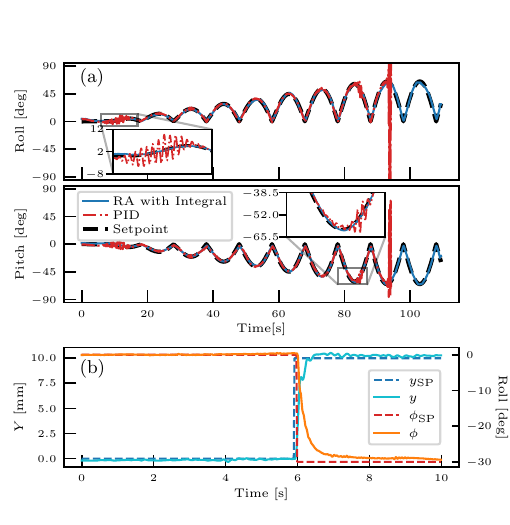}%
    }%
    \vspace{-4mm}
    \caption{Trajectory tracking experiments.
    Orientation is represented as XYZ intrinsic Euler angles: roll $\stateRoll$, pitch $\statePitch$, and yaw $\stateYaw$; $\stateYaw$ is the uncontrollable rotation mode.
    \textit{Left:} Trajectory tracking experiments with simultenous position and orientation setpoint changes without integrators (colormap indicates condition number) using Object I in the OctoMag. Gyroscopic coupling from imperfections in the levitator leads to large drifts in $\stateYaw$.
    \textit{Right:} (a) Benchmarking the reduced-attitude controller (with integral action) against an equivalent PID baseline for large-angle tracking up to 65$^\circ$ using Object III in the 13-coil system. The PID controller suffers from self-amplifying nutation oscillations due to uncompensated 3D cross-coupling, destabilizing at $t\approx\qty{93}{\second}$. The proposed nonlinear controller remains robust and stable at large spatial rotations. (b) Simultaneous step response of Object II in the OctoMag. The relative overshoots are 5.2\% for $y$ and 2.0\% for $\phi$, with steady-state errors (with integral control) of \qty{0.26}{\milli\meter} and 0.69$^\circ$, respectively.
    }
    \label{fig:trajectory_tracking_figures}
\end{figure*}
\begin{figure}[!t]
    \centering
    \includegraphics[width=0.95\linewidth, trim=30 6 12 45, clip]{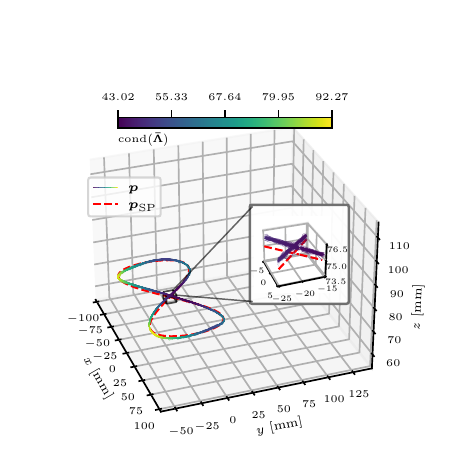}
    \caption{3D tracking of Object III along an xy figure-eight trajectory using a 13-coil system with integrators enabled (colormap indicates condition number). Over the $\approx$28-minute experiment (84 cycles), position RMS errors remained below 0.6 mm (x: 0.31 mm, y: 0.43 mm, z: 0.59 mm) with periodic repeatability (average standard deviations < 0.13 mm for position, $0.95^{\circ}$ for roll, 1.34$^{\circ}$ for pitch). Thermal camera monitoring confirmed the coil surface temperature stayed below 35$^{\circ}$C throughout the duration.}
    \label{fig:3d_lissajous_tracking_integrators_enabled}
    \vspace{-1.5em}
\end{figure}
\section{Results and Discussion}\label{sec:results_and_discussion}

In this section, we evaluate our analysis through trajectory tracking experiments on the physical setup described in \cref{subsec:experimental_setup__sec_experimental_setup_and_results}.

Controllers are implemented in Python on Ubuntu 20.04 LTS, running on a 20-core 3.4 GHz Intel i7-14700HK CPU, and 64 GB of RAM.
We use Robot Operating System (ROS) as the middleware for communication between software components.
To reduce delays, parts of the code are just-in-time compiled and cached with Numba \cite{lamNumbaLLVMbasedPython2015}. The total delay, from marker position acquisition to current setpoint arrival at the coil drivers, is approximately \qty{4}{\milli\second} (sensing and communication $\approx$\qty{3.8}{\milli\second}, computation $\approx$\qty{0.3}{\milli\second}, and driver response $\approx$\qty{0.1}{\milli\second}). To evaluate the system's sensitivity to latency, we artificially injected delays into the control loop; an additional \qtyrange{3}{5}{\milli\second} significantly degraded control performance, highlighting the critical importance of our low-latency software architecture.
Velocity estimates are obtained using first order backward finite differences.

We conducted a series of experiments across the OctoMag and the 13-coil system using three distinct objects to demonstrate the object- and platform-agnostic capabilities of our control pipeline.
\cref{fig:trajectory_tracking_figures} (left) shows simultaneous position and orientation setpoint tracking without integrator action using Object I in the OctoMag. Imperfections in the object such as a non-diagonal inertia tensor lead to gyroscopic coupling in the dynamics of $\stateAngvelBfZ, \stateAngvelBfZDot$ and cause the yaw angle $\stateYaw$ to drift. Notably, our control strategy successfully preserves stable levitation even when subjected to a manually induced yaw spin of 5 rotations per second (see video attachment).
With integrators enabled, \cref{fig:trajectory_tracking_figures} (right) highlights the controller's dynamic responsiveness using simultaneous step responses (Object II) and demonstrates its superiority over the PID baseline during large-angle tracking (Object III in the 13-coil system), which destabilizes due to uncompensated 3D cross-coupling.
Additionally, a continuous 28-minute 3D figure-eight maneuver (\cref{fig:3d_lissajous_tracking_integrators_enabled}) confirms long-term stability and high repeatability.
Furthermore, we conducted additional experiments to assess the controller's robustness against unknown mass (up to 16\%) and magnetization mismatches ($\pm$5\%); these results, alongside further trajectory tracking experiments, are showcased in the supplementary video.

\section{Conclusion}\label{sec:conclusion}
This paper demonstrated stabilization and trajectory tracking of a levitating object with a permanent magnetic dipole using an eMNS. To highlight the object- and platform-agnostic capabilities of our control pipeline, we utilized both the \octomag{} and a custom 13-coil eMNS as our actuation platforms, testing across three distinct levitators.
With this work, we aim to demonstrate the dynamic capabilities of eMNS - an important step toward clinical use cases, which are currently dominated by quasi-static models and feedforward control.

We modeled the magnetic field  generated by the \octomag{}'s coil currents using MPEM, a parametric field model designed for computational efficiency.
A stable feedback controller was synthesized to regulate the levitator's position and reduced attitude. 
Key design factors for achieving a robust performance, such as the choice of the levitator's mass, dipole moment, and the eMNS's bandwidth frequency were discussed.
Notably, our controller demonstrated clear superiority over an equivalent baseline PID formulation by reliably tracking large angular deflections up to 65$^\circ$, a capability not previously reported in the literature on magnetic levitation.

Our nonlinear reduced-attitude controller is currently constrained by the motion-capture system, which cannot track full rotations. Future work will extend the state estimation to handle full rotations and fully exploit the controller's performance. Another promising direction is to investigate whether levitation can be achieved on clinically ready, human-scale electromagnetic navigation systems such as Navion \cite{gervasoni2024navion}. Our long-term goal is to push the limits of eMNS systems and achieve highly complex, precise maneuvers that may enable new medical applications.


\section*{Acknowledgment}

The authors would like to thank Prof. Peter Berkelman from the University of Hawaii at Manoa for sharing valuable insights based on his years of experience in developing magnetic levitation systems.
We also thank Felix Grüninger and Thomas Steinbrenner from MPI-IS Tübingen for their support with electronic and hardware designs essential to this project. Additionally, the authors utilized ChatGPT and Gemini for sentence restructuring to improve the readability of the manuscript.
Finally, we thank the team at MagnebotiX AG for technical support with the \octomag{} eMNS.

\section*{Conflict of Interest}
Bradley Nelson is a co-founder of MagnebotiX AG, which commercializes the \octomag{} system.
The other authors declare no conflict of interest.

\bibliographystyle{ieee/bib/IEEEtran}
\bibliography{bibliography,nonsync_bibliography}

\end{document}